# Applying Genetic Algorithm for Prioritization of Test Case Scenarios Derived from UML Diagrams

Sangeeta Sabharwal, Ritu Sibal and Chayanika Sharma

Department of computer Science and IT

Netaji Subhas Institute of Technology

Delhi, India

**Abstract**

Software testing involves identifying the test cases which discover errors in the program. However, exhaustive testing of software is very time consuming. In this paper, a technique is proposed to prioritize test case scenarios by identifying the critical path clusters using genetic algorithm. The test case scenarios are derived from the UML activity diagram and state chart diagram. The testing efficiency is optimized by applying the genetic algorithm on the test data. The information flow metric is adopted in this work for calculating the information flow complexity associated with each node of the activity diagram and state chart diagram. If the software requirements change, the software needs to be modified and this requires re – testing of the software. Hence, to take care of requirements change, a stack based approach for assigning weights to the nodes of activity diagram and state chart diagram has also been proposed. In this paper, we have extended our previous work of generating test case scenarios from activity diagram by also considering the concurrent activities in nested activity diagram.

*Keywords: software testing, genetic algorithm, activity diagram, state chart diagram, CFG, SDG, test case*

## 1. INTRODUCTION

Software testing is one of the major and primary techniques for achieving high quality software. Software testing is done to detect presence of faults, which cause software failure. However, software testing is a time consuming and expensive task [1], [3]. It consumes almost 50% of the software system development resources [2], [3], [9]. Testing can be done either manually or automatically by use of testing tools. It is found that automated software testing is better than manual testing. However, very few test data generation tools are commercially available today [4].

Evolutionary testing is an emerging methodology for automatically producing high quality test data [11]. Genetic algorithms (GA) are well known form of the evolutionary algorithms conceived by John Holland in United States during late sixties [10], [12]. GA has been applied in many optimization problems for generating test plans for functionality testing, feasible test cases and in many other areas [6], [7]. GA has also been used in model based test case generation [2], [5]. Various techniques have been proposed for generating test data/test cases automatically using GA in structural testing [3], [4]. GA has also been applied in the regression testing, object oriented unit testing as well as in the black box testing for the automatic generation of the test cases [5], [6], [11].

Unified modelling language (UML) is the de-facto standard for modelling object – oriented software systems. UML provides diagrams to represent the static as well as the dynamic behaviour of a system [16]. Class, component and deployment diagrams are used to represent the static behaviour of the system whereas activity, sequence and state diagrams are used to represent the dynamic behaviour. UML Activity diagram shows the activities of the object, so the operations can be realized in the design stage itself [17].

In this paper, we have proposed a technique for prioritization of test case scenarios derived from activity diagram and state chart diagram of UML using the concept of basic information flow (IF) metric, stack and GA. The paper presents the extended work of our previous work [18]. In this paper, as extension to our previous work we propose a technique for generating test cases from state chart diagram and nested activity diagrams wherein concurrent activities are also occurring. The paper is divided into 6 sections. Section 2 describes the basic structure of GA. In section 3, our proposed approach is discussed while section 4 and section 5 describe the test case scenarios derived from nested activity diagram and state chart diagram. Section 6 concludes the paper and gives an overview of our future work.

## 2. GENETIC ALGORITHM

In the past, evolutionary algorithms have been applied in many real life problems. GA is one such evolutionary algorithm. GA has emerged as a practical, robust optimization technique and search method. A GA is a search algorithm that is inspired by the way nature evolves species using natural selection of the fittest individuals.

The possible solutions to the problem being solved are represented by a population of chromosomes. A chromosome is a string of binary digits and each digit that makes up a chromosome is called a gene. This initial population can be totally random or can be created manually using processes such as greedy algorithm. The pseudo code of a basic algorithm for GA is as follows [18]:-

```
Initialize (population)
Evaluate (population)
While (stopping condition not satisfied)
  {
    Selection (population)
    Crossover (population)
    Mutate (population)
    Evaluate (population)

  }
```

A GA uses three operators on its population which are described below:-

- **Selection**: A selection scheme is applied to determine how individuals are chosen for mating based on their fitness. Fitness can be defined as a capability of an individual to survive and reproduce in an environment. Selection generates the new population from the old one, thus starting a new generation. Each chromosome is evaluated in present generation to determine its fitness value. This fitness value is used to select the better chromosomes from the population for the next generation.

- **Crossover or Recombination**: After selection, the crossover operation is applied to the selected chromosomes. It involves swapping of genes or sequence of bits in the string between two individuals. This process is repeated with different parent individuals until the next generation has enough individuals. After crossover, the mutation operator is applied to a randomly selected subset of the population.

- **Mutation**: Mutation alters chromosomes in small ways to introduce new good traits. It is applied to bring diversity in the population.

## 3. PROPOSED APPROACH

This section illustrates the details of our proposed approach for test case prioritization using GA. The approach uses nested activity diagram and state chart diagram for deriving the test case scenarios. Activity diagram depicts the functional view of the system by modelling the flow of control from one activity to another. An activity represents an operation that results in the change of the state in the system. In activity diagram, the nodes represent activities. The main constructs used in the activity diagrams are shown in Fig.1.

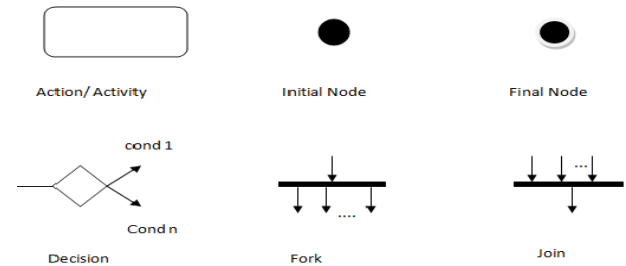

Fig. 1 Main Constructs used in Activity Diagram

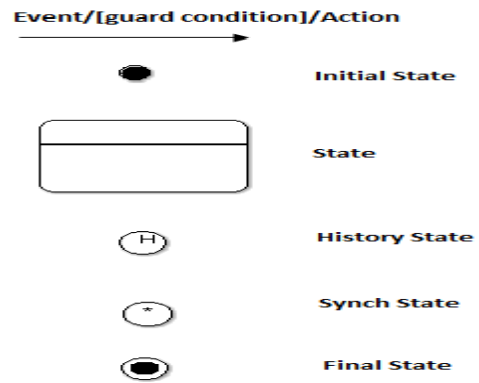

Fig. 2 Main Constructs used in State Chart Diagram

For the purpose of test case generation, the activity diagram is converted into a control flow graph (CFG) where each node represents an activity and the edges of the flow graph depict the control flow of the activities.

In our work, we have also derived the test cases form the state chart diagram. The main constructs used in the state chart diagram are shown in Fig.2. The event with guard condition and action generates a new state. The state chart diagram is converted into an intermediate graph called as State Dependency Graph (SDG) for test case generation.

Path testing involves generating a set of paths from CFG or SDG that will cover every branch in the program and finding the set of test case scenarios that will traverse every activity or state in these scenarios. It may be very tedious, expensive and time consuming to achieve this goal due to various reasons. For example, there can exist infinite paths when a CFG or SDG have loops. In our approach we propose to find the critical path that must be tested first using the concept of IF metric, stack and GA. While finding the path we are ensuring that every loop is traversed at most once.

### 3.1 Procedure

If requirements change, the software needs to be redesigned to comply with the software specification before the implementation phase. During requirement engineering, addition or deletion of user requirements causes changes in the activity diagram as well as state chart diagram where the action or states are directly derived from user requirements. If right changes are not taken care during the requirement phase then they are carried over to the implementation phase where effort, time and resources are unnecessarily wasted in correcting the user requirements change.

We take care of the issue of requirements change by prioritizing the nodes of CFG and SDG using the stack based memory allocation approach and IF metrics. In the stack based memory allocation approach, the data or info is pushed or popped only at one end called top of stack. The stack uses last in first out (LIFO) approach. Node pushed first is removed last from the stack. The top is incremented when node is inserted and decremented when node is deleted.

The stack based memory allocation approach is explained with an illustration in section 4 and section 5 respectively. Thus the steps involved in identifying the critical path clusters for an activity and state chart diagram are as follows:-

a) Convert the activity diagram into CFG and state chart diagram into SDG respectively.

b) Assign the weights to the nodes of CFG or SDG as the case may be, by using stack based weight assignment approach and the Basic IF Model.

In the proposed stack based memory allocation technique, for the nodes of the CFG or SDG each node is assigned a weight, w based on number of operations to access element in the stack. But to access or modify the node (data), we have to pop all the data above it. Higher the number of operations required to access the node, higher is the weight and hence the complexity of the node. If the weight of the node or number of operations to access the node increases the cost of modifying the node also increases.

In the Basic IF model [15], information flow metrics are applied to the components of system design. In our work, the component is taken as a node in the CFG and SDG. The IF is calculated for each node of a CFG and SDG. For example, the IF of node A i.e. IF (A) is calculated using equation given below:-

$$IF(A) = [FANIN(A) \times FANOUT(A)] \quad (1)$$

Where FANIN (A) is the count of the number of other nodes that can call, or pass control to node A and FANOUT (A) is the number of nodes that are called by node A. The IF is calculated for each node of a CFG and SDG. The weighted nodes in the path are summed together and the complexity of each path is calculated.

Therefore, the sum of the weight of a node by stack based weight assignment approach and IF complexity contributes to the total weight of a node of CFG and SDG.

c) Selection: - The decision nodes of the CFG and SDG form the chromosome. A chromosome or test data will therefore be a binary string where a single bit or multiple bits in a string will correspond to a decision node of the graph. Number of bits in the chromosome will depend upon the number of decision nodes and the type of graph being used. For example, in CFG if there are four decision nodes, a four bit binary string will form a chromosome or an individual in the population. The fitness value of each chromosome is calculated by applying the stack based weight assignment approach and Basic IF model. The chromosomes with high fitness value are selected as the parents for the reproduction. The fitness value of each chromosome is calculated by using the formula given below:-

$$F = \sum_{i=1}^{n} w_i \quad (2)$$

Where, $w_i$ is weight of $i^{th}$ node in a path under consideration and n is number of nodes in a current path. Weight of $i^{th}$ node is the sum of IF complexity and stack based complexity given by equation given below.

$$w_i = IF(i) + STACKBASEDWEIGHT(i) \quad (3)$$

d) Crossover: - There are number of techniques of crossover, but all require swapping of genes or sequence of bits in the chromosome. It involves swapping between two individuals or test data in our case. In our work, we assume the probability of crossover is taken as 80% [3]. The random number r is generated from 0 to 1. Crossover is done if r < 0.8

**ALGORITHM 1**

1. Convert the activity diagram into CFG.

2. Use the decision nodes to generate the test data or chromosome population randomly.

3. for each test data $i = 1....n$,

   a) Traverse the CFG by applying Depth first search (DFS) as well as Breadth first search (BFS) and identify the paths.

      (i) Find the neighbour node for the current node having next higher depth $d_i$ by applying DFS.

      (ii) For each concurrent node, $c_i$ traverse the next neighbour node having next higher breadth value, $b_i$.

      (iii) Update the top pointer, size $s$ and $k$ of the Stack.

      (iv) Assign weight, $w$ to each node by applying weight assignment algorithm.

   b) Calculate the fitness value of each test data by using equation (2) and stack based weight assignment approach.

   c) For sub activity of each node calculate the fitness value using 80-20 rule and by using equation (2).

   d) Select initial test data by ranking the fitness of the chromosomes.

   e) If initial population is not enough randomly generate them.

        If r < 0.8, perform crossover
        Else if r < 0.2, perform mutation
      end if
     end if
   end

4. If test data for all the paths have not been covered, then repeat the GA process.

5. Else end system

Fig. 3 Proposed Algorithm for generating test scenarios from the activity diagram

condition is satisfied. This follows from our assumption that crossover probability should be 80%.

e) Mutation:-Mutation is done to introduce new traits or bring diversity in the population to avoid the local optima. In mutation the bits are flipped from 0 to 1 and vice versa. In our work, we assume the probability of mutation as 20%. The random number is generated from 0 to 1. If r < 0.2 condition is satisfied, then the bits of the test data are mutated randomly.

The algorithm for our proposed approach to generate test scenarios for activity diagram and state chart diagram is shown in Fig.3 and Fig.5 respectively while the weight assignment algorithm is shown in Fig.4.

**ALGORITHM 2**

1. for each activity node $a_i$, $i = 1...n$,

   a) Push nodes of CFG on the stack using DFS and BFS approach.

   b) Determine the maximum size, $s_{max}$ of the stack.

   c) for $i = 1$ to $s_{max}$, assign $w = s_{max} - k$ to each node of the CFG, where $s_{max}$ is maximum size of stack and k is number of nodes above current node.

   d) for each decision node, $d_i$

      (i) Assign the same weight, $w$ to branching nodes.
      (ii) Insert the next neighbour nodes of branching nodes and update top.
      (iii) Neglect the branching nodes and decision nodes that have been inserted previously.
     end

   e) Assign same weight, $w$ to concurrent nodes.
   end

Fig. 4 Algorithm for the weight assignment

## 4. TEST CASE SCENARIOS DERIVED FROM ACTIVITY DIAGRAM

In this section we will explain our approach by taking the activity diagram of a shipping order system (Fig.6). The activity diagram is converted into CFG as shown in Fig.7. In this case study we have taken nested activity diagram that also takes into account concurrency among action nodes.

---

**ALGORITHM 3**

1. Convert the state chart diagram into SDG.

2. Use the fork nodes in SDG to generate the test data or chromosome population randomly.

3. for each test data $i = 1 .... n$,

   a) Traverse the SDG by applying Depth first search (DFS).

     (i) Find the neighbour node for the current node having next higher depth $d$ by applying DFS.

     (ii) Update the top pointer, size $s$ and $k$ of the stack, where $k$ is number of nodes above the current node.

     (iii) Assign weight, $w$ to the node by applying the weight assignment algorithm.

   b) Calculate the fitness value of each test data by using equation (2) and stack based weight assignment approach.

   c) Select initial test data by ranking the fitness of the chromosomes.

   d) If initial population is not enough randomly generate them.

         If $r < 0.8$, perform crossover
         Else if $r < 0.2$, perform mutation
       end if
     end if
   end

4. If test data for all the paths have not been covered, then repeat the GA process.

5. Else end system

Fig. 5 Proposed Algorithm for generating test scenarios from state chart diagram

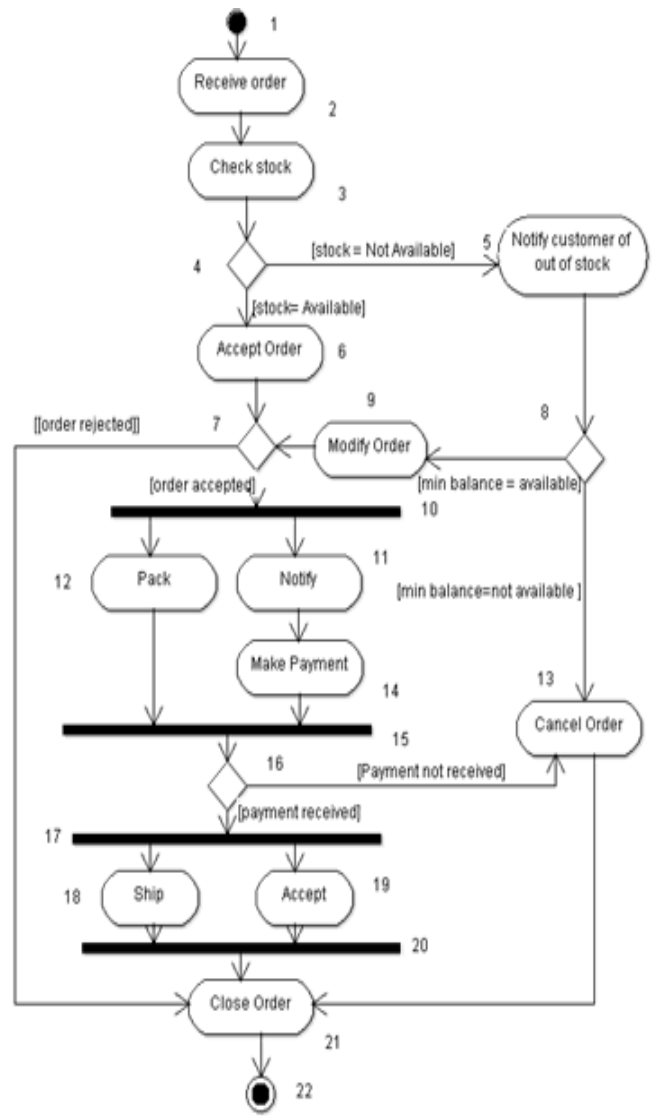

Fig. 6 Activity diagram of shipping order system

In Fig.7, the fork node 10 shows concurrency between nodes 11 and 12 whereas the fork node 17 shows concurrency between node 18 and 19 nodes respectively. The concurrent nodes are shown as rounded rectangles in Fig.7.

For Fig.7, the stack based weight assignment to all nodes of CFG is shown in Fig.8. The maximum size of stack is $s_{max} = 18$. The initial node of CFG i.e. node 1 is pushed first hence the weight, w for node 1 is $s_{max} - k$ i.e. $18 - 0 = 18$ where $s_{max}$ is maximum size of stack and k is number of nodes before the lowest node i.e. number of nodes before node 1. Similarly the priority for node 2 is 17 and node 3 is 16. The complexity of node 1 is highest i.e. 18. Therefore, making changes to node 1 will require the highest number of operations to access it.

The sum of stack based priority number (A) and the IF complexity of each node (B) is equal to the total complexity of each node.

In this example, we have taken the nested activity for modify order. The nested activity diagram includes the fork and merges nodes. The Fig.9 shows the nested sub activity diagram of Modify Order shown in Fig.6. The Fig.10 has one fork and merge node showing the concurrency among the nodes. The corresponding CFG for Fig.9 is shown in Fig.10.

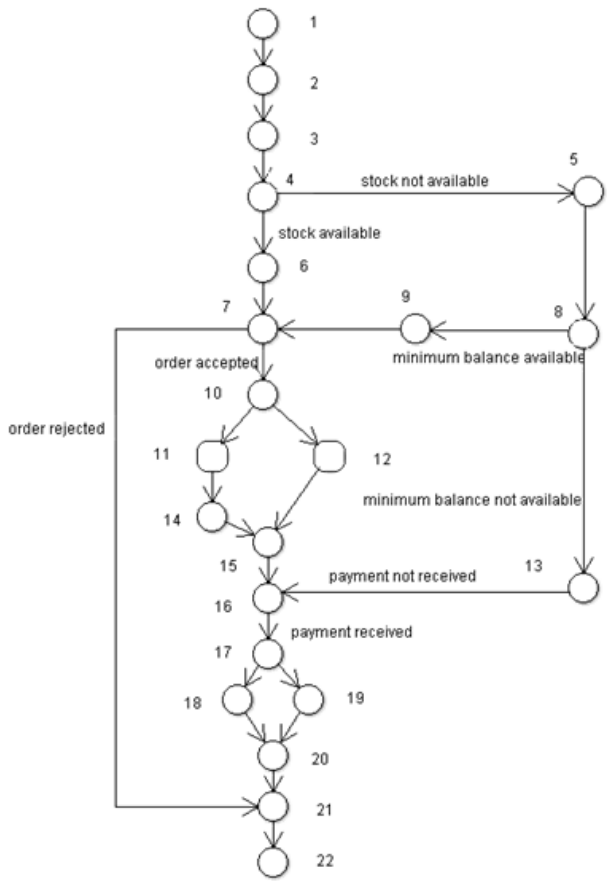

Fig. 7 CFG of Activity diagram of shipping order system

Now as clear from Fig. 7, there are four decision nodes namely 4, 7, 8 and 16 which will form the chromosome in our case. Therefore, corresponding to the chromosome value 0000 the decision nodes i.e. 4, 7, 8 and 16 will evaluate to false condition. The path to be followed using this chromosome value is 1, 2, 3, 4, 5, 8, 13, 21, 22. The IF complexity of each node is calculated shown in Table I where, FI is Fan In and FO is Fan Out of the corresponding node. The fitness value of test data 0000 is $18 + 18 + 17 + 17 + 15 + 15 + 19 + 32 + 22 = 173$.

| Nodes | | K | Size, s | Weight, w = $s_{max} - k$ |
|---|---|---|---|---|
| 22 | | 17 | 18 | 18-17=1 |
| 21 | | 16 | 17 | 18-16=2 |
| 20 | | 15 | 16 | 18-15=3 |
| 22 | 19 | 14 | 15 | 18-14=4 |
| 21 | 18 | 13 | 14 | 18-13=5 |
| 13 or 17 | | 12 | 13 | 18-12=6 |
| 16 | | 11 | 12 | 18-11=7 |
| 15 | | 10 | 11 | 18-10=8 |
| 14 | | 9 | 10 | 18-9=9 |
| 11 | 22 | 8 | 9 | 18-8=10 |
| 12 or 22 | 21 | 7 | 8 | 18-7=11 |
| 10 or 21 | 9 or 13 | 6 | 7 | 18-6=12 |
| 7 | 8 | 5 | 6 | 18-5=13 |
| 5 or 6 | | 4 | 5 | 18-4=14 |
| 4 | | 3 | 4 | 18-3=15 |
| 3 | | 2 | 3 | 18-2=16 |
| 2 | | 1 | 2 | 18-1=17 |
| 1 | | 0 | 1 | 18-0=18 |

Fig. 8 Stack based weight assignments to nodes of CFG for the shipping order

Table 1: Complexity of nodes of the CFG of shipping order

| Node | Complexity based on pop operation, (A) | IF= Fan In (a) * Fan Out (a), (B) | Total Complexity = (A+B) |
|---|---|---|---|
| 1 | 18 | 0 | 18 |
| 2 | 17 | 1 | 18 |
| 3 | 16 | 1 | 17 |
| 4 | 15 | 2 | 17 |
| 5 | 14 | 1 | 15 |
| 6 | 14 | 1 | 15 |
| 7 | 13 | 2 | 15 |
| 8 | 13 | 2 | 15 |
| 9 | 12 | 1 | 13 |
| 10 | 12 | 2 | 14 |
| 11 | 10 | 2 | 12 |
| 12 | 11 | 1 | 12 |
| 13 | 12+6 | 1 | 19 |
| 14 | 9 | 1 | 10 |
| 15 | 8 | 2 | 10 |
| 16 | 7 | 2 | 9 |
| 17 | 6 | 2 | 8 |
| 18 | 5 | 1 | 6 |
| 19 | 4 | 1 | 5 |
| 20 | 3 | 2 | 5 |
| 21 | 12+11+5+2 | 2 | 32 |
| 22 | 11+10+1 | 0 | 22 |

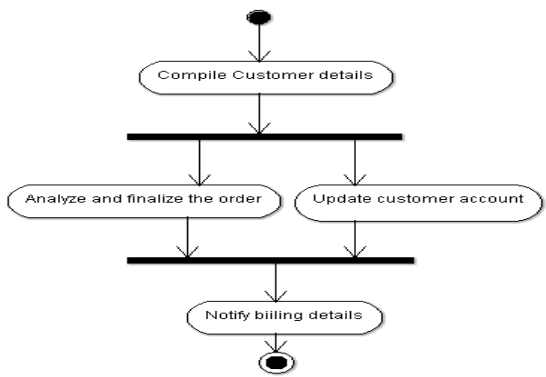

Fig. 9 Nested Activity diagram of Modify order

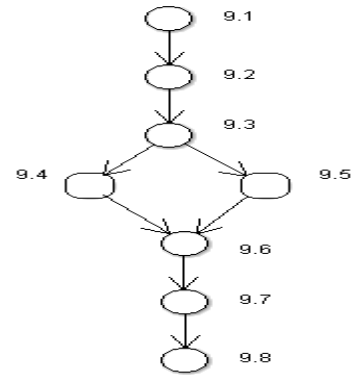

Fig. 10 CFG of modify order (nested activity)

Top

| Nodes | K | Size, s | Weight = $s_{max}-k$ |
|---|---|---|---|
| 9.8 | 7 | 8 | 8 - 7 = 1 |
| 9.7 | 6 | 7 | 8 - 6 = 2 |
| 9.6 | 5 | 6 | 8 - 5 = 3 |
| 9.5 | 4 | 5 | 8 - 4 = 4 |
| 9.4 | 3 | 4 | 8 - 3 = 5 |
| 9.3 | 2 | 3 | 8 - 2 = 6 |
| 9.2 | 1 | 2 | 8 - 1 = 7 |
| 9.1 | 0 | 1 | 8 - 0 = 8 |

Fig. 11 Stack based weight assignment to nodes of nested activity diagram (Fig.10)

Table 2: Complexity of CFG (Fig.10)

| Node | Complexity based on pop operation, (A) | IF= Fan In (a) * Fan Out (a), (B) | Total Complexity = (A+B) |
|---|---|---|---|
| 9.1 | 8 | 0 | 8 |
| 9.2 | 7 | 1 | 8 |
| 9.3 | 6 | 2 | 8 |
| 9.4 | 5 | 1 | 6 |
| 9.5 | 4 | 1 | 5 |
| 9.6 | 3 | 2 | 5 |
| 9.7 | 2 | 1 | 3 |
| 9.8 | 1 | 0 | 1 |

The total complexity for the nested activity of modify order is the summation of the total complexity of each node i.e. 8 + 8 + 8 + 6 + 5 + 5 + 3 + 1 = 44. The total complexity of activity Modify Order i.e. node 9 (Fig.6) = complexity of node 9 + complexity of nested activity of node 9 i.e. 13 + 44 = 57. Hence, the total complexity of node 9 is 57.

We start the process by randomly generating the initial population as shown in Table 3.

Initial Population: - 0011, 0001, 1100 and 1111, where X is the test data, F(X) is the fitness value, r is random number generated from 0 to 1. F'(X) is new computed fitness value after crossover and mutation operation. C is crossover and M is mutation operation. In our example, GA is run for 12 iterations. The test data 0011 will follow the path 1, 2, 3, 4, 5, 8, 9, 7, 21, 22 and the corresponding fitness value for the test data is 18 + 18 + 17 + 17 + 15 + 15 + 57 + 15 + 32 + 22 = 226. The test data 0001 will follow the path 1, 2, 3, 4, 5, 8, 13, 21, 22 and the corresponding fitness value for the test data is 8 + 18 + 17 + 17 + 15 + 15 + 19 + 32 + 22= 173. The test data 1100 will follow the path 1, 2, 3, 4, 6, 7, 10, 11, 12, 14, 15, 16, 13, 21, 22 and the corresponding fitness value is 18 + 18 + 17 + 17 + 15 + 15 + 14 + 12 + 12 + 10 + 10 + 9 + 19 + 32 + 22 = 240. The test data 1111 will follow the path 1, 2, 3, 4, 6, 7, 10, 11, 12, 14, 15, 16, 17, 18, 19, 20, 21, 22 and the corresponding fitness value for the test data is 18 + 18 + 17 + 17 + 15 + 15 + 14 + 12 + 12 + 10 + 10 + 9 + 8 + 6 + 5 + 5 + 32 + 22 = 245.

Table 3: Iteration 1

| S. No. | X | F(X) | R | C | M | F'(X) |
|---|---|---|---|---|---|---|
| 1. | 0011 | 226 | 0.327 | 0111 | 0111 | 317 |
| 2. | 0001 | 173 | 0.867 | 0001 | 0001 | 173 |
| 3. | 1100 | 240 | 0.912 | 1100 | 1100 | 240 |
| 4. | 1111 | 245 | 0.490 | 1011 | 1011 | 154 |

Table 4: Iteration 2

| S. No. | X | F(X) | R | C | M | F'(X) |
|---|---|---|---|---|---|---|
| 1. | 0111 | 317 | 0.372 | 0101 | 0101 | 173 |
| 2. | 1100 | 240 | 0.541 | 1110 | 1110 | 240 |
| 3. | 0001 | 173 | 0.934 | 0001 | 0001 | 173 |
| 4. | 1011 | 154 | 0.860 | 1011 | 1011 | 154 |

Table 5: Iteration 3

| S. No. | X | F(X) | R | C | M | F'(X) |
|---|---|---|---|---|---|---|
| 1. | 1100 | 240 | 0.841 | 1110 | 1110 | 240 |
| 2. | 0101 | 173 | 0.415 | 0001 | 0001 | 173 |
| 3. | 0001 | 173 | 0.301 | 0101 | 0101 | 173 |
| 4. | 1011 | 154 | 0.971 | 1011 | 1011 | 154 |

.
.
.
.

Table 6: Iteration 11

| S. No. | X | F(X) | R | C | M | F'(X) |
|---|---|---|---|---|---|---|
| 1. | 0111 | 317 | 0.712 | 0111 | 0111 | 317 |
| 2. | 0111 | 317 | 0.431 | 0111 | 0111 | 317 |
| 3. | 0111 | 317 | 0.692 | 0111 | 0111 | 317 |
| 4. | 1111 | 245 | 0.169 | 1111 | 0111 | 317 |

Table 7: Iteration 12

| S. No. | X | F(X) | R | C | M | F'(X) |
|---|---|---|---|---|---|---|
| 1. | 0111 | 317 | 0.436 | 0111 | 0111 | 317 |
| 2. | 0111 | 317 | 0.946 | 0111 | 0111 | 317 |
| 3. | 0111 | 317 | 0.871 | 0111 | 0111 | 317 |
| 4. | 0111 | 317 | 0.192 | 0111 | 0011 | 226 |

After the 12$^{th}$ iteration as shown in table 7, the test data 0111 have the highest fitness value i.e. 317. So, the path corresponding to this chromosome value i.e. 1, 2, 3, 4, 5, 8, 9, 7, 10, 11, 12, 14, 15, 16, 17, 18, 19, 20, 21, 22 should be the one which must be tested first. Next we will discuss an example illustrating our approach to generate and prioritize test case scenarios from state chart diagram.

## 5. TEST CASE SCENARIOS DERIVED FROM STATE CHART DIAGRAM

In our next case study we have applied our approach on the state chart diagram of the student enrolment system as shown in Fig.12. In student enrolment system, student enrolled is an event, seat available is a guard condition and add student is an action. If the guard condition becomes true then action is performed and open for enrolment state is generated. The intermediate graph is called as SDG (Fig.13).

In Fig.13, INIT represents start process, PR represents proposed state, SC represents scheduled state, OE represents open for enrolment state, FU represents Full state and CE represents closed to enrolment state respectively. The events are shown as e1, e2, e3.......e12 where e1 is scheduled, e2 is registration open, e3 is student apply for enrolment, e4 is enrolment closed, e5 is student apply for waiting list, e6 is close enrolment, e7 is student dropped, e10 is seat allotment and e8, e9, e11, e12 are cancelled events respectively. As shown in Fig.12, there are four decision nodes namely 2, 3, 4, and 6 respectively which will form the chromosomes in our case study.

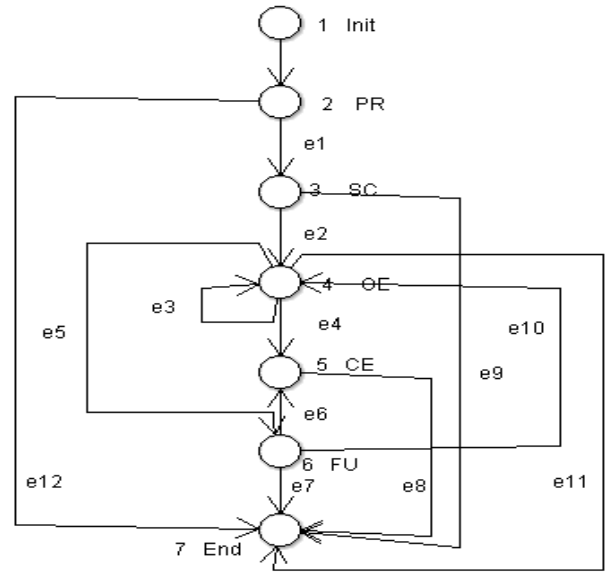

Fig. 13 SDG of student enrolment system

| Nodes | K | Size, s | Weight = $s_{max}$-k |
|---|---|---|---|
| 7 | 5 | 6 | 6-5 = 1 |
| 5 or 6 or 7 | 4 | 5 | 6-4 = 2 |
| 4 | 3 | 4 | 6-3 = 3 |
| 3 or 7 | 2 | 3 | 6-2 = 4 |
| 2 | 1 | 2 | 6-1 = 5 |
| 1 | 0 | 1 | 6-0 = 6 |

Fig. 14 Stack based weight assignment to nodes of SDG

The events e1, e2, e3, e4, e5, e6, e7, e8, e9, e10, e11 and e12 represent the edges and the states represent the corresponding nodes in the SDG. The test data corresponding to decision nodes 2, 3, 4 and 6 are shown in Table 8. These test data show the events representing edges in the SDG and are part of test case scenarios. For example test data involving events e1, e2, e5 and e7 will lead to path consisting of nodes 1, 2, 3, 4, 6 and 7. It is to be noted that in Table 8 we have considered all the test data leading to valid as well as invalid paths.

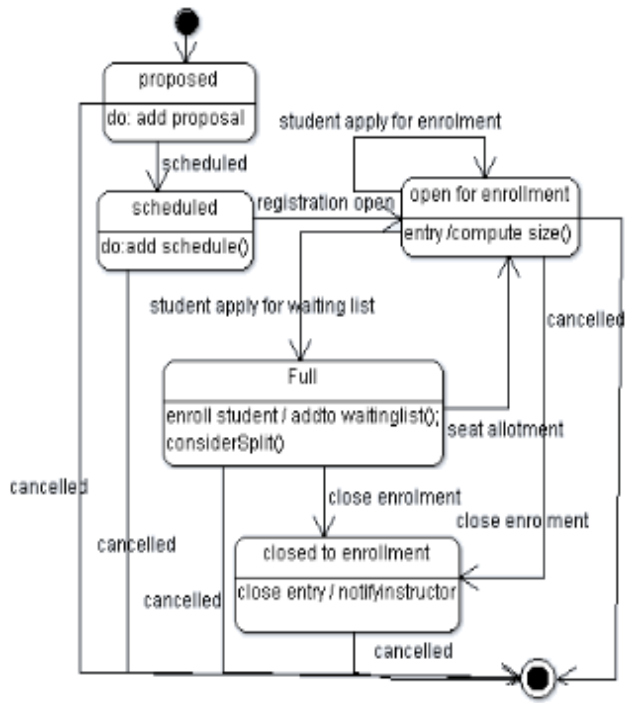

Fig. 12 State chart diagram of student enrolment system

Table 8: Test cases for SDG

|       | Decision Nodes |     |     |     |
|-------|-----|-----|-----|-----|
| S.No. | 2   | 3   | 4   | 6   |
| 1.    | e1  | e2  | e4  | e7  |
| 2.    | e1  | e2  | e4  | e10 |
| 3.    | e1  | e2  | e5  | e7  |
| 4.    | e1  | e2  | e5  | e10 |
| 5.    | e1  | e2  | e11 | e7  |
| 6.    | e1  | e2  | e11 | e10 |
| 7.    | e1  | e9  | e4  | e7  |
| 8.    | e1  | e9  | e4  | e10 |
| 9.    | e1  | e9  | e5  | e7  |
| 10.   | e1  | e9  | e5  | e10 |
| 11.   | e1  | e9  | e11 | e7  |
| 12.   | e1  | e9  | e11 | e10 |
| 13.   | e12 | e2  | e4  | e7  |
| 14.   | e12 | e2  | e4  | e10 |
| 15.   | e12 | e2  | e5  | e7  |
| 16.   | e12 | e2  | e5  | e10 |
| 17.   | e12 | e2  | e11 | e7  |
| 18.   | e12 | e2  | e11 | e10 |
| 19.   | e12 | e9  | e4  | e7  |
| 20.   | e12 | e9  | e4  | e10 |
| 21.   | e12 | e9  | e5  | e7  |
| 22.   | e12 | e9  | e5  | e10 |
| 23.   | e12 | e9  | e11 | e7  |
| 24.   | e12 | e9  | e11 | e10 |
| 25.   | e1  | e2  | e3  | e7  |
| 26.   | e1  | e2  | e3  | e10 |
| 27.   | e1  | e9  | e3  | e7  |
| 28.   | e1  | e9  | e3  | e10 |
| 29.   | e12 | e2  | e3  | e7  |
| 30.   | e12 | e2  | e3  | e10 |
| 31.   | e12 | e9  | e3  | e7  |
| 32.   | e12 | e9  | e3  | e10 |

In this example as clear from Fig.13 in a state, maximum four events can take place. Therefore each event in SDG is represented in two bit format. In our case, the events e1 and e12 at node 2 are represented by 00 and 01, events e2 and e9 at node 3 are represented by chromosome 00 and 01, events e11, e5, e4 and e3 at node 4 are represented by 00, 01, 10 and 11 respectively. Events e7 and e10 at node 6 are represented by 00 and 01 respectively. Therefore, each event in SDG is represented in two bit format. As there are four decision nodes in the SDG, the chromosome here will consist of a binary string of 8 bits. In our proposed approach, loops are traversed at most once. Therefore, node having self loop is considered only once in a path. For example, the test case scenario involving events e1, e2, e3, e7 will follow the path consisting of nodes 1, 2, 3, 4, 5 and 7 .The node 4 containing self loop is traversed only once in the path and the next node to be followed after loop is the nearest neighbour having shortest distance i.e. next node to be followed after node 4 is 5 as the distance between node 4 and node 5 is small as compared to node 6 and 7.

The sum of stack based priority number (A) and the IF complexity of each node (B) is equal to the total complexity of each node. The IF complexity of each node of SDG is shown in Table 9 where, FI is Fan In and FO is Fan Out for the corresponding node. The test data representation involving decision nodes is shown in Fig.15 using two bit format for each decision node.

```
  e1        e2        e4        e7

| 0  0  |  0  0  |  1  0  |  0  0  |   Test case 1

  e1        e2        e4        e10

| 0  0  |  0  0  |  1  0  |  0  1  |   Test case 2
```

•

•

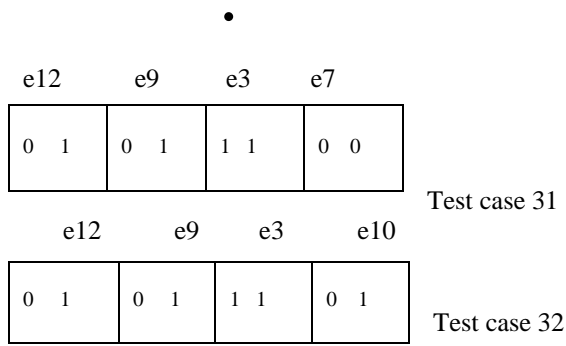

Fig.15 Test case representation using two bit format

Table 9: Complexity of nodes of SDG

| Node | Complexity based on pop operation, (A) | Fan In(a) * Fan Out(a), (B) | Total Complexity = (A+B) |
|---|---|---|---|
| 1 | 6 | 0 | 6 |
| 2 | 5 | 2 | 7 |
| 3 | 4 | 2 | 6 |
| 4 | 3 | 6 | 9 |
| 5 | 2 | 2 | 4 |
| 6 | 2 | 3 | 5 |
| 7 | 4+2+1=7 | 0 | 7 |

We start the process by randomly generating the initial population as shown in Table 10.

Initial Population: - 00011000, 01001000, 01000101, 00000100 where X is the test data, F(X) is the fitness value, r is random number generated from 0 to 1. F'(X) is new computed fitness value after crossover and mutation operation. C is crossover and M is mutation operation.

The test data 00011000 will traverse the edges e1, e9, e4, e7 and hence will follow the path 1, 2, 3 and 7 and the corresponding fitness value of test data is 6 + 7 + 6 + 7 = 26. The test data 01001000 will traverse the edges e12, e2, e4 and e7 and hence will follow the path 1, 2 and 7 and the corresponding fitness value is 6 + 7 + 7 = 20. The test data 01000101 will follow the edges e12, e2, e5, e10 and will follow the path 1, 2 and 7 and the corresponding fitness value is 6 + 7 + 7 = 20. Similarly the test data 00000100 will follow the edges e1, e2, e5, e7 and hence will follow the path 1, 2, 3, 4, 6 and 7 and the fitness value computed is 6 + 7 + 6 + 9 + 5 + 7 = 40.

Table 10: Iteration 1

| S. No | X | F(X) | R | C | M | F'(X) |
|---|---|---|---|---|---|---|
| 1. | 00011000 | 26 | 0.829 | 00011000 | 00011000 | 26 |
| 2. | 01001000 | 20 | 0.876 | 01001000 | 01001000 | 20 |
| 3. | 01000101 | 20 | 0.712 | 01000100 | 01000100 | 20 |
| 4. | 00000100 | 40 | 0.686 | 00000101 | 00000101 | 54 |

Table 11: Iteration 2

| S. No | X | F(X) | R | C | M | F'(X) |
|---|---|---|---|---|---|---|
| 1. | 00011000 | 26 | 0.917 | 00110000 | 00110000 | 26 |
| 2. | 01001000 | 20 | 0.126 | 01001000 | 01000100 | 20 |
| 3. | 01000100 | 20 | 0.814 | 01000100 | 01000100 | 20 |
| 4. | 00000100 | 54 | 0.562 | 00000100 | 00000100 | 54 |

Table 12: Iteration 3

| S. No. | X | F(X) | R | C | M | F'(X) |
|---|---|---|---|---|---|---|
| 1. | 00110000 | 26 | 0.833 | 00110000 | 00110000 | 26 |
| 2. | 01000100 | 20 | 0.415 | 01000100 | 01000100 | 20 |
| 3. | 01000100 | 20 | 0.124 | 01000100 | 01001000 | 20 |
| 4. | 00000100 | 54 | 0.612 | 00000100 | 00000100 | 54 |

.
.
.

Table 13: Iteration 8

| S.No. | X | F(X) | R | C | M | F'(X) |
|---|---|---|---|---|---|---|
| 1. | 01000100 | 20 | 0.841 | 01000100 | 00100000 | 20 |
| 2. | 00000101 | 54 | 0.912 | 00000101 | 00000000 | 54 |
| 3. | 00001001 | 39 | 0.104 | 00001001 | 00000101 | 54 |
| 4. | 00000101 | 54 | 0.809 | 00000101 | 00010101 | 54 |

Table 14: Iteration 9

| S.No. | X | F(X) | R | C | M | F'(X) |
|---|---|---|---|---|---|---|
| 1. | 01000100 | 20 | 0.971 | 01000100 | 01000100 | 20 |
| 2. | 00000101 | 54 | 0.501 | 00000101 | 00000101 | 54 |
| 3. | 00001001 | 54 | 0.412 | 00000101 | 00000101 | 54 |
| 4. | 00000101 | 54 | 0.972 | 00000101 | 00000101 | 54 |

In our example, after the 9$^{th}$ iteration as shown in Table 14, the test data e1, e2, e5, e10 have the highest fitness value i.e. 54. So, the path corresponding to the chromosome 00000101 should be the one which must be tested first. Therefore the path consisting of nodes 1, 2, 3, 4, 6, 4, 5 and 7 consisting of events e1, e2, e5 and e10 should be the one that will be tested first.

# 6. CONCLUSION AND FUTURE WORK

In this paper a GA based approach is proposed for identifying the test path that must be tested first. Test paths or scenarios are derived from activity diagram and state chart diagram respectively. The proposed approach makes use of IF model and GA to find the path to be tested first. Our future work involves applying the proposed approach on other UML diagrams like sequence diagram and using this technique for white box testing and object oriented testing. A tool is also being developed to support this proposed approach.